\documentclass[a4paper,10pt,twoside]{cpc-hepnp}
\usepackage{pifont}
\usepackage{multicol}
\usepackage{graphicx}
\usepackage{booktabs}
\usepackage{amssymb,bm,mathrsfs,bbm,amscd}
\usepackage[tbtags]{amsmath}
\usepackage{lastpage}
\usepackage{CJK}

\begin{document}
\begin{CJK*}{GBK}{song}

\fancyhead[co]{\footnotesize Lingling Ma et al: Geometry and optics
calibration for WFCTA telescopes using star light }

\footnotetext[0]{Received XX March XXXX}

\title{Geometry and optics calibration of WFCTA prototype telescopes using star light
\thanks{Supported by 100 Talents Programme of The Chinese Academy of Sciences,
Knowledge Innovation Program of The Chinese Academy of Sciences ( H85451D0U2 ),
National Natural Science Foundation of China ( 10975145 ).} }
\author{%
      MA Ling-Ling$^{1;1)}$\email{llma@ihep.ac.cn}%
    \quad Bai Yun-Xiang$^{1)}$
    \quad CAO Zhen$^{1)}$
    \quad Chen Ming-Jun$^{1)}$\\
    \quad Chen Li-Hong$^{1)}$
     \quad Chen Song-Zhan$^{1)}$
     \quad Chen Yao$^{1)}$
     \quad Ding Kai-Qi$^{1)}$\\
     \quad He Hui-Hai$^{1)}$
      \quad Liu Jia$^{1)}$
      \quad Liu Jia-Li$^{1)}$
      \quad Li Xiao-Xiao$^{1)}$\\
      \quad Ma Xin-Hua$^{1)}$
      \quad Sheng Xiang-Dong$^{1)}$
       \quad Xiao Gang$^{1)}$
       \quad Zha Min$^{1)}$\\
      \quad Zhang Shou-Shan$^{1)}$
      \quad Zhang Yong$^{1)}$
     \quad Zhao Jing$^{1)}$
     \quad Zhou Bin$^{1)}$
}

\maketitle

\address{%
1~Institute of High Energy Physics, Chinese Academy of Sciences,
Beijing 100049, China
 }

\begin{abstract}
The Large High Altitude Air Shower Observatory project is proposed
to study high energy gamma ray astronomy ( 40 GeV-1 PeV ) and cosmic
ray physics ( 20 TeV-1 EeV ). The wide field of view Cherenkov
telescope array, as a component of the LHAASO project, will be used
to study energy spectrum and compositions of cosmic ray by measuring
the total Cherenkov light generated by air showers and shower
maximum depth. Two prototype telescopes have been in operation since
2008. The pointing accuracy of each telescope is crucial to the
direction reconstruction of the primary particles. On the other hand
the primary energy reconstruction relies on the shape of the
Cherenkov image on the camera and the unrecorded photons due to the
imperfect connections between photomultiplier tubes. UV bright stars
are used as point-like objects to calibrate the pointing and to
study the optical properties of the camera, the spot size and the
fractions of unrecorded photons in the insensitive areas of the
camera.
\end{abstract}



\begin{multicols}{2}

\section{Introduction}
The Large High Altitude Air Shower Observatory ( LHAASO
)~\citep{LHAASO} project aims to study 40 GeV-1 PeV gamma ray
astronomy and 20 TeV-1 EeV cosmic ray physics at Yangbajing ( 4300 m
a.s.l. ), Tibet, China, near the AS$_{\gamma}$ and ARGO-YBJ
experiments. The Wide Field of View ( FOV ) Cherenkov telescope
Array ( WFCTA ), as a component of the LHAASO project, is designed
to study cosmic ray energy spectrum specie by specie by measuring
the energy and $X_{max}$ depth of each air shower.

Two WFCTA prototype telescopes have been constructed and placed
nearby the ARGO-YBJ experimental hall. The two telescopes can be
operated in both monocular and stereo modes, while coincident
observation with the ARGO-YBJ detector is achieved off-line.

Each telescope is made up of two main parts, the reflector and the
camera. The reflector consists of 20 spherical mirrors with a radius
curvature $R$ of 4740$\pm$20 mm, corresponding to a total area of
4.7 m$^{2}$. The reflecting efficiency of the mirrors is about 82\%
for light with wavelength larger than 300 nm. A camera is placed at
the focal plane which is 2305 mm away from the reflector center to
optimize the spot shape of a point-like object. The camera is
composed of 256 flat hexagonal photomultipliers tubes ( PMTs ) each
of which has a diameter of 40 mm, corresponding to a FOV of about
$1^{\circ}\times1^{\circ}$. The PMTs are arranged in 16 columns and
16 rows forming a total FOV of
$14^{\circ}\times16^{\circ}$~\citep{He-2007}. The maximum quantum
efficiency of PMTs can reach $30\%$ at 420 nm~\citep{PMT}. The
signals of the PMTs are digitized by 50 MHz Flash Analog to Digital
Converters ( FADCs ). The whole system is hosted in a shipping
container with a dimension of 2.5 m$\times$2.3 m$\times$3 m. The
container is mounted on a standard dump-truck frame with a hydraulic
lift that allows the container to be tilted in any elevation angles
from 0 to 60 degrees. The pointing direction of the telescope can be
easily changed~\citep{He-2007}.

The pointing accuracy and geometry properties of the telescopes are
crucial to the arrival direction and $X_{max}$ reconstructions. The
energy reconstruction relies on the recorded Cherenkov image,
however, due to the imperfect physical junction between PMTs, some
Cherenkov photons are unrecorded in the joints. In order to improve
the accuracy of energy reconstruction the optical properties are
studied.

In this paper, we describe a method to calibrate the geometry and
optical properties by using UV bright stars. While monitoring the UV
Cherenkov light from air showers, the telescopes are also sensitive
to the UV light from stars crossing the FOV of the telescopes. With
their well known positions, orders of magnitude more accurate than
the required resolution of WFCTA, and their point-like shape, the
stars are ideal tools to test the pointing direction of each
telescope. Using stars the optical properties of the telescope are
studied.

\section{Night sky background and star signal}
In addition to record the Cherenkov light from air showers, the
camera also records the night sky background light (NSB). When a
bright star enters the FOV of the telescope, light from the star is
added to the diffused NSB. The changes of the recorded NSB reflect
the light sources crossing the FOV of the PMTs and the weather
conditions. A star stays about 4 minutes in the FOV of a PMT, during
this time light from the star is added to the diffused NSB. While
NSB change due to weather change behaves differently, which usually
lasts much longer than 4 minutes and almost all PMTs are affected at
the same time. This enables star signals to be discriminated from
weather change.

For each air shower event, the signal of Cherenkov light only lasts
a few nanoseconds, while the trigger window lasts $18\mu s$, so
telescopes record NSB in most of the trigger window. In order to
reduce the fluctuation the recorded NSB is averaged in every $10s$.
Fig.~\ref{nsb} shows a typical NSB curve in one night recorded by a
PMT. After subtracting the diffused background, peaks due to stars
are clearly seen. The peak amplitude of a star light curve in a PMT
depends on its UV brightness and its projected position to the PMT
on the camera. In a typical night, many stars can be seen by each
PMT.

\begin{center}
\includegraphics[width=9cm]{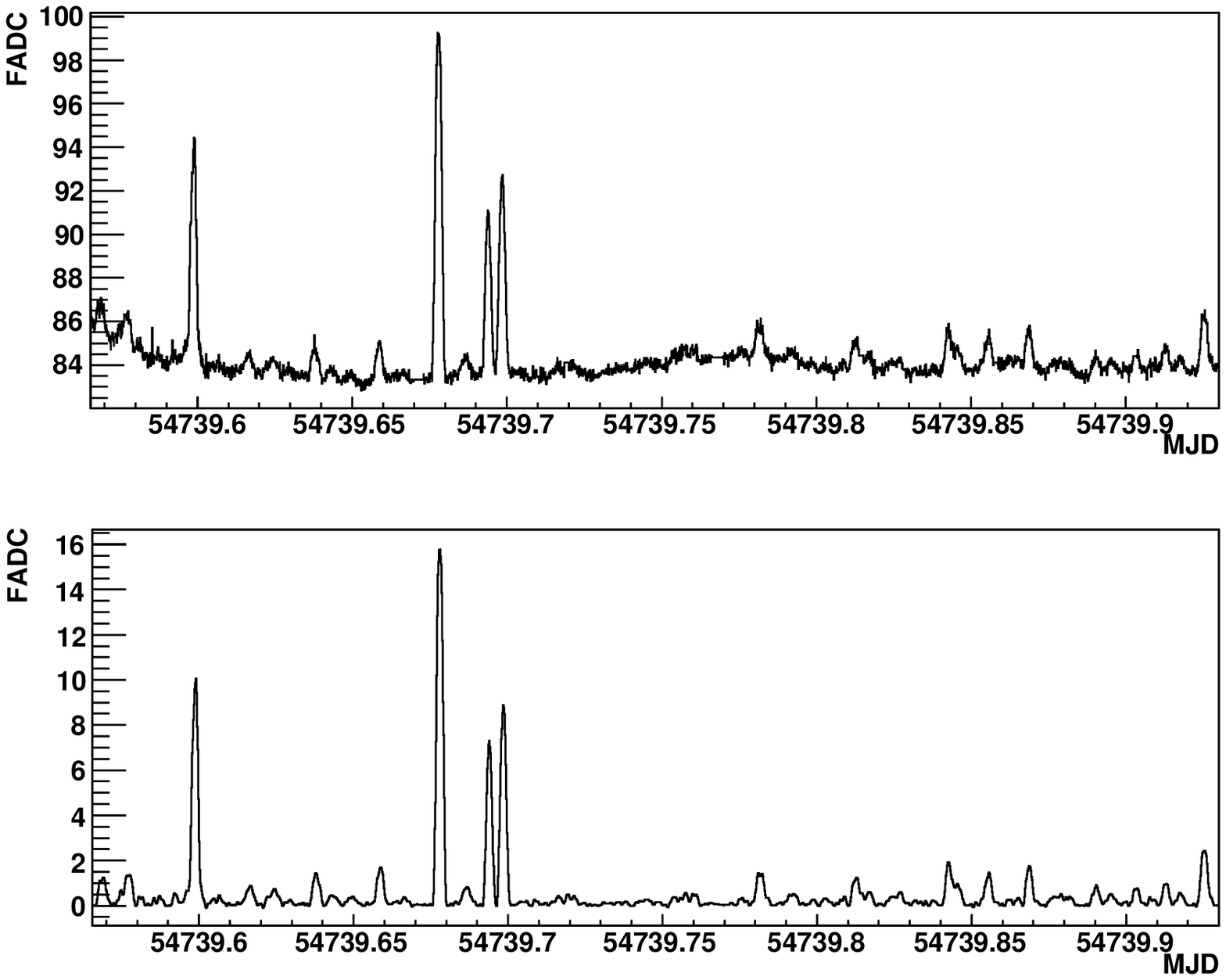}
\figcaption{\label{nsb}   A typical NSB curve recorded by a PMT in
one night before (upper) and after (bottom) subtracting the diffused
NSB. }
\end{center}
\section{Pointing direction}
Stars with well known positions and brightness are used as guiders
of the telescopes. In our analysis, the TD 1 catalog is used, which
has four different wavelength bands, 1565, 1965, 2365 and 2740 $\AA$
respectively~\citep{TD1}. Since the WFCTA telescopes are sensitive
in the near UV band, stars with flux in the 2740${\AA}$ band above
$1\times10^{-11}erg/cm^{2}/s$
 are used.

The pointing direction of each telescope can be changed through the
container which encloses the whole telescope system. Only very rough
pointing direction (about 200$^{\circ}$ in azimuth and 60$^{\circ}$
in elevation) is known through it. Using the rough pointing
direction, the time when the brightest star appears in the FOV of
the telescope can be found through the brightest PMT. However, due
to the large size of the PMT, the position of the PMT can not show
accurately the position of the brightest star on the camera. So the
weighted center position ($x_{0}$, $y_{0}$) of the PMT and its
neighbors within $2^{\circ}$ is considered as the position of the
star on the camera. When the star is in the middle of the camera in
the horizontal direction, the star has the same azimuth angle with
the telescope, while the elevation angle of the star is equal to the
elevation of telescope plus the distance between the position of the
star and the center of the camera. The obtained pointing direction
of the telescope is more accurate than the one from the container.

After getting the pointing direction of the telescope through the
brightest star, orphan stars which have no surrounding stars within
$2^{\circ}$ are used to correct the pointing direction of the
telescope by using iteration method.

The accuracy of the pointing direction can be described by
differences between positions of stars in the local coordinates and
the obtained ones. Fig.~\ref{pointing} shows the distribution of the
differences for one telescope. An accuracy better than
$0.05^{\circ}$ is obtained in less than 20 minutes with five stars
in the FOV.
\begin{center}
\includegraphics[width=8cm]{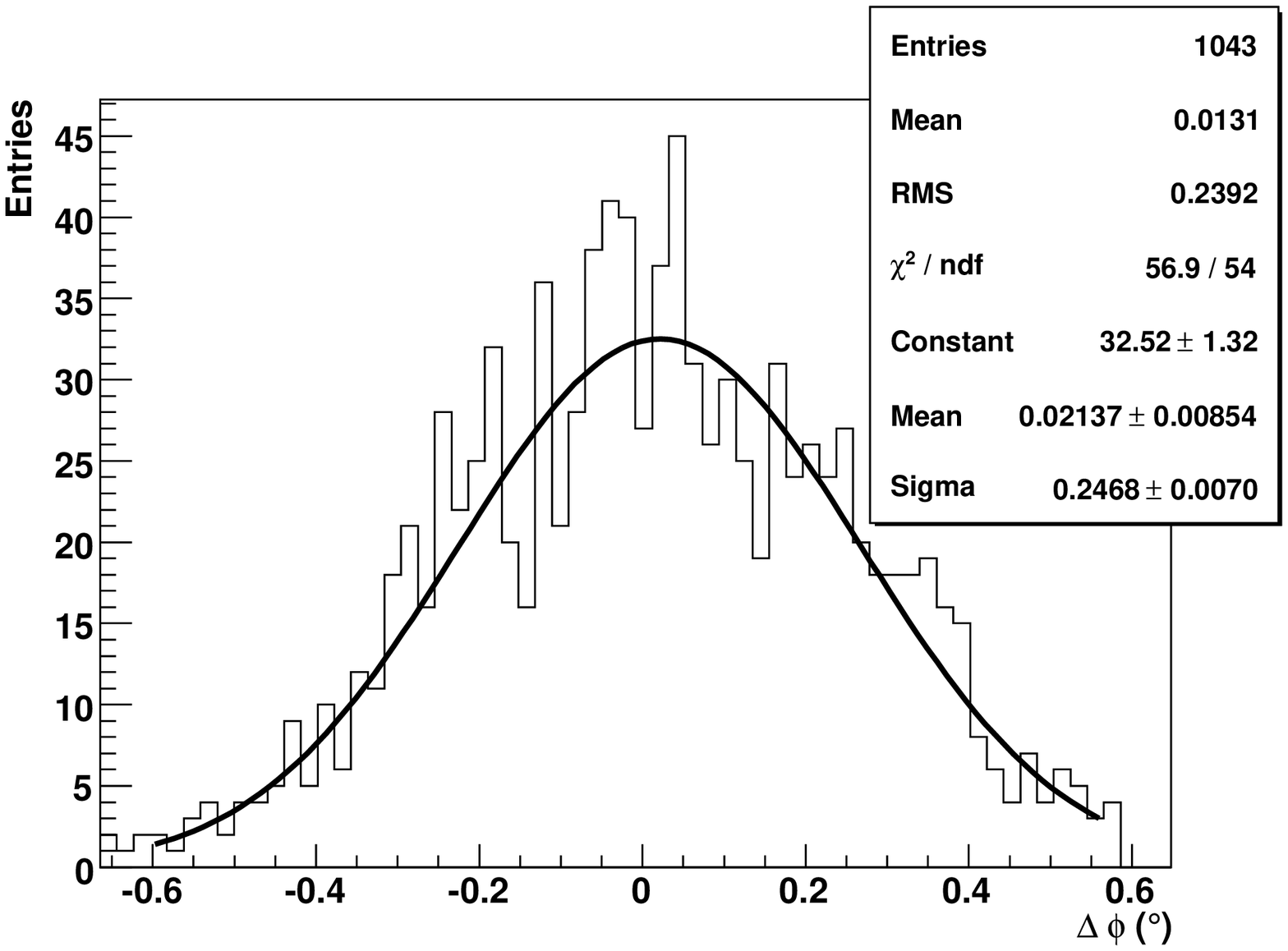}
\includegraphics[width=8cm]{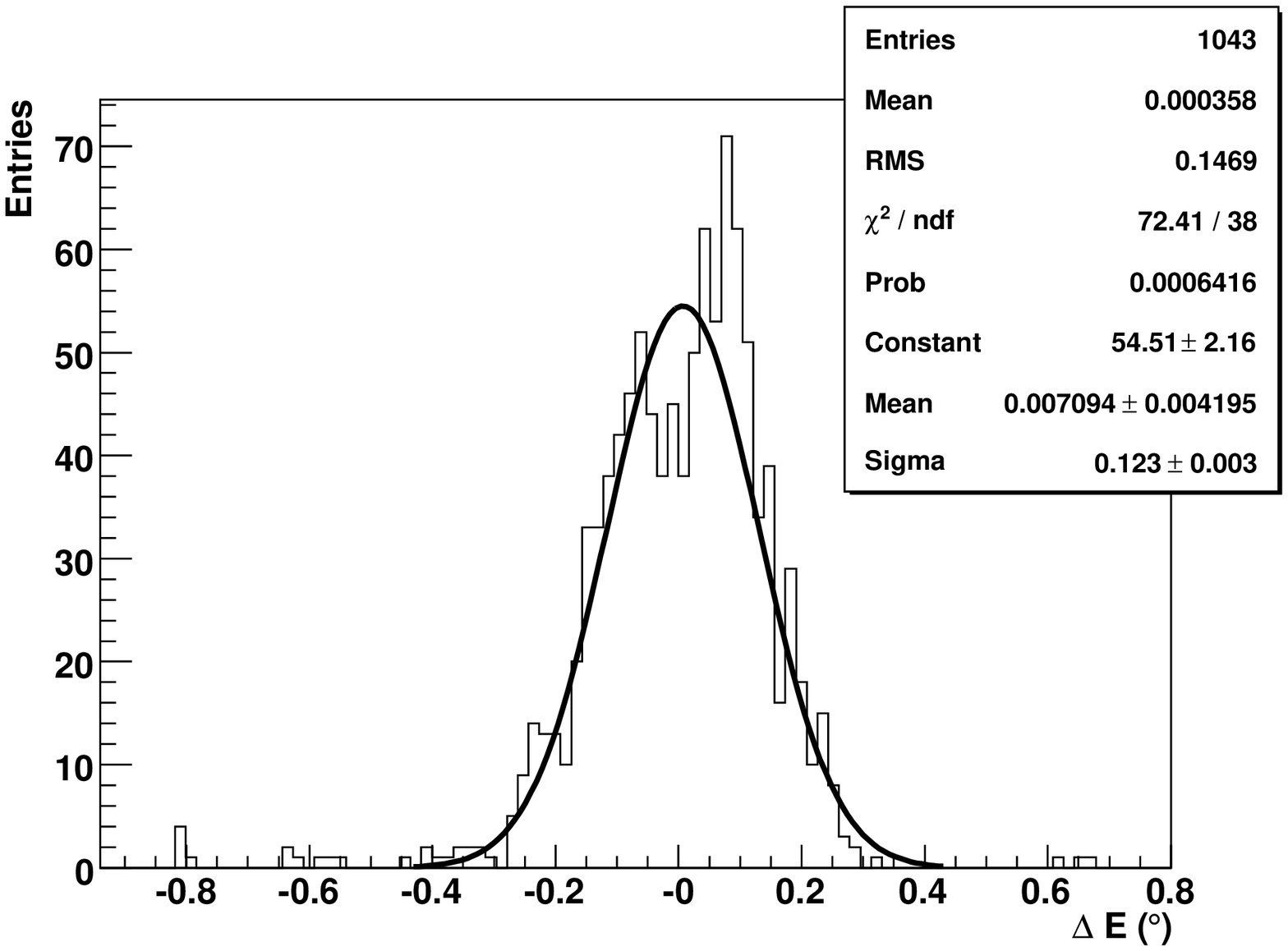}
\figcaption{\label{pointing} The distributions of the differences
between positions of stars and the obtained ones in azimuth (upper)
and elevation (bottom). A Gaussian fit is superposed to each
distribution.}
\end{center}

Fig.~\ref{PointingMonitor} shows the pointing of one telescope
changing with time in two observation periods, i.e. two months. Both
azimuth and elevation are very stable. The big change of the
elevation was due to lowering and elevating the container between
the two periods.

\begin{center}
\includegraphics[width=8cm]{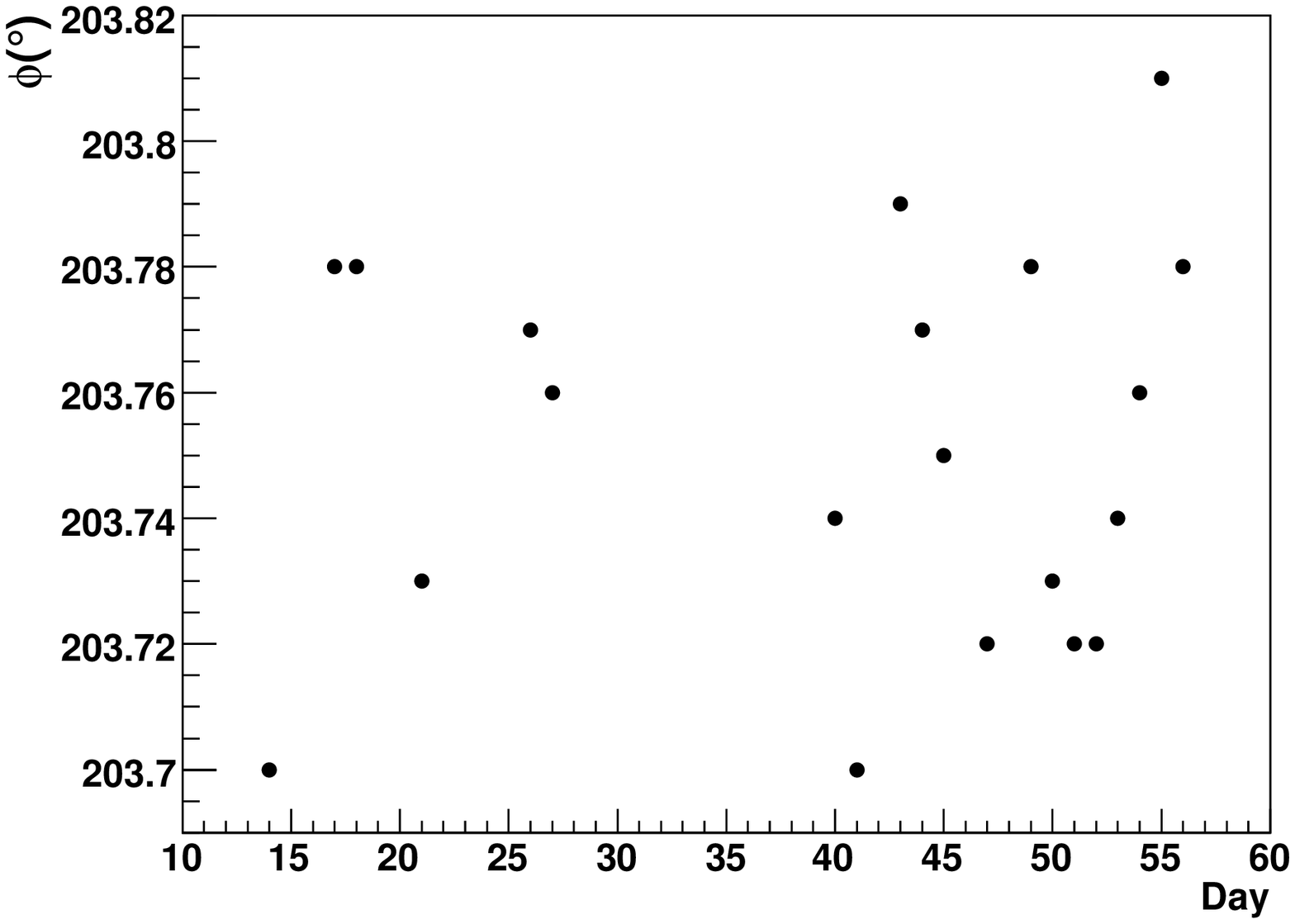}
\includegraphics[width=8cm]{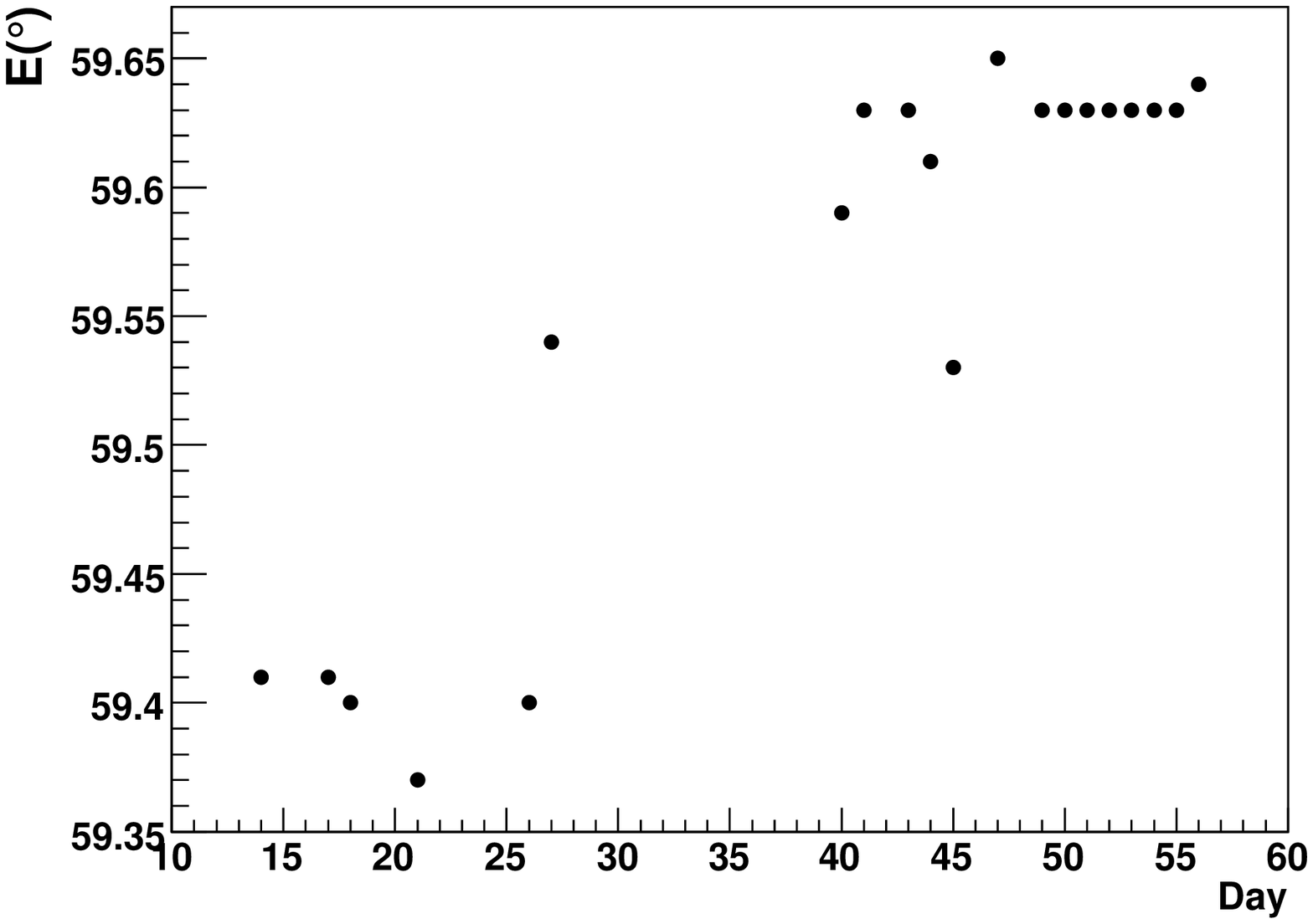}
\figcaption{\label{PointingMonitor} The pointing directions of one
telescope versus time in two months (Up: azimuth, Bottom:
Elevation). Each point represents one day.}
\end{center}

\section{Camera geometry}

The camera geometry calibration is done after pointing correction of
each telescope. The calibration includes the following four
parameters. The first one ( $P_{1}$ ) is a scaling of the tubes away
from the center of the tube cluster. To first order $P_{1}$ corrects
for deviations in the radius curvature of a mirror and for changes
in the effective camera-mirror distance due to the treatment of the
flat camera as a curved surface. The second parameter ( $P_{2}$ )
describes the rotation angle of the camera around the mirror axis.
The last two parameters ( $P_{3}$, $P_{4}$) indicate the offsets of
the shift in the position of the entire camera with respect to the
mirror axis. After the four parameters corrections the positions
($x_{c,t}$, $y_{c, t}$) of stars on the camera at time $t$ are
modified by Eq.~(\ref{eq2}) and Eq.~(\ref{eq3}).
\begin{equation}
\label{eq2} x_{c, t}'=(1+p_{1})(x_{c, t}cos(p_{2})-y_{c,
t}sin(p_{2}))+p3
\end{equation}
\begin{equation}
\label{eq3} y_{c, t}'=(1+p_{1})(x_{c, t}sin(p_{2})+y_{c,
t}cos(p_{2}))+p4
\end{equation}
The $x_{c, t}'$ and $y_{c, t}'$ in Eq.~(\ref{eq2}) and
Eq.~(\ref{eq3}) are the stars' coordinates after geometry
correction. The four parameters can be obtained by the least squared
method. The $\chi^{2}$ is shown in Eq.~(\ref{eq1}).
\begin{equation}
\label{eq1}
\chi^{2}=\sum_{star}\sum_{t}\frac{(x_{c,t}'-x_{star,t})^{2}+(y_{c,t}'-y_{star,t})^{2}}{\sigma_{x}^{2}+\sigma_{y}^{2}}
\end{equation}

In Eq.~(\ref{eq1}), the $x_{star, t}$, $y_{star, t}$ are the stars'
exact positions on the camera at time $t$. The $\sigma_{x}$ and
$\sigma_{y}$ are the RMS of the $x_{c, t}'$ and $y_{c, t}'$,
respectively. By minimizing the $\chi^{2}$, the values of $P_{1}$,
$P_{2}$, $P_{3}$ and $P_{4}$ are found to be $-1.5\%$,
$0.6^{\circ}$, $-0.05^{\circ}$ and
 $-0.05^{\circ}$. These parameters are used in the detector
 simulation and data reconstruction.

\section{Spot size and variations of the number of observed photons}

Photons from any given direction form a quasi-Gaussian-shaped spot
on the camera, which is infected by imperfection of the mirrors
surface. The spot size depends on the angular distance to the
optical axis. The larger the angular distance to the optical axis,
the larger the spot size and the more it deviates from a Gaussian
shape due to coma of the image. The spot size as an important
parameter that affects the Cherenkov images of air showers is taken
into account in the ray tracing procedure in both data analysis and
detector simulation.
\begin{center}
\includegraphics[width=9cm]{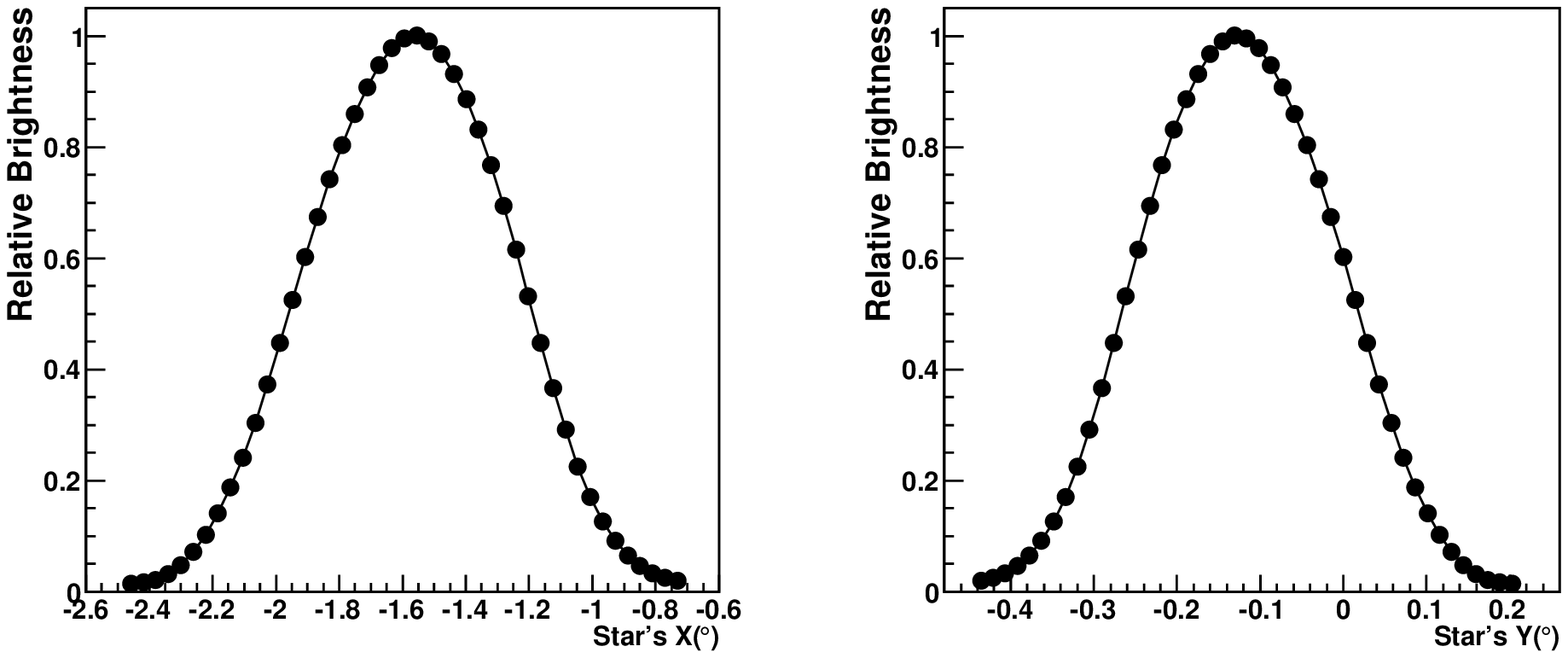}
\figcaption{\label{spot_xy} One example of light curves in X (left)
and Y (right) directions when a star passes through a PMT's FOV.}
\end{center}

Bright stars can be considered as perfect point sources. The light
from a bright star is collected by mirrors and projected to the
camera, forming a light spot. The camera records the light spot in
poor resolution due to the large pixel size. If a PMT is on the
track of a bright star, the PMT gets brighter and brighter as the
star goes nearer, dimmer and dimmer as the star leaves. If the spot
size is much smaller than the track length of the star in the FOV of
the PMT, the light curve recorded by the PMT will be
rectangular-shaped, with the width approximately equal to the track
length, while if the spot size is much larger, the light curve also
will be rectangular-shaped, with the width equal to the spot size.
In our case, the spot size is similar to the pixel size, the light
curve becomes quasi-Gaussian-shaped.

Fig.~\ref{spot_xy} shows one example. The spot size can be estimated
by fitting the light curve accordingly. In order to get rid of the
affects of nearby stars, only orphan stars, which have no nearby
stars within $2^{\circ}$, are used. Fig.~\ref{spot_fit} shows the
variation of the spot size with angular distance to mirror axis. The
spots become larger from the center of the camera to the edge. The
spot size obtained from bright stars is used to improve the energy
and arrival direction reconstruction.

\begin{center}
\includegraphics[width=8cm]{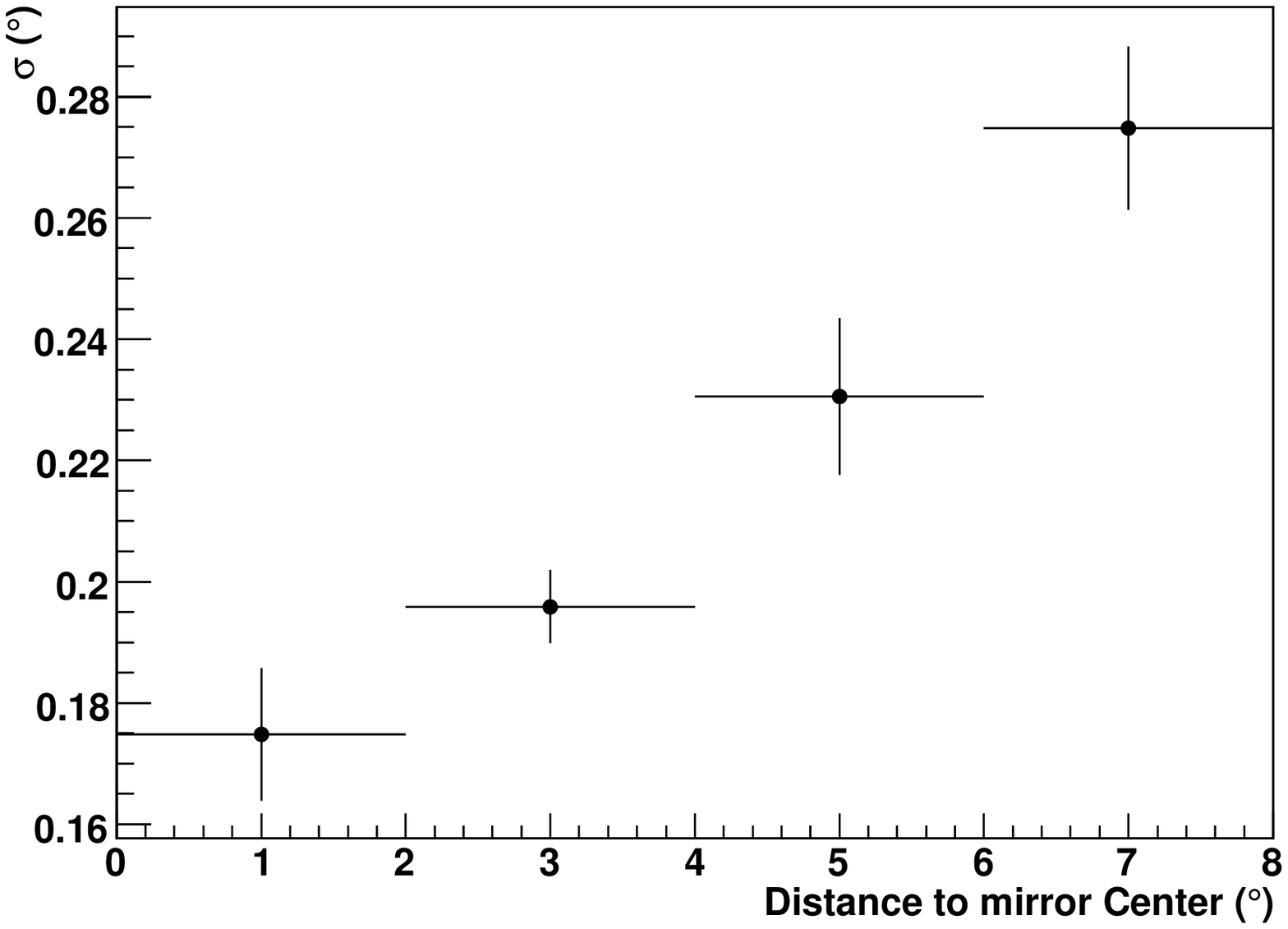}
\figcaption{\label{spot_fit} Spot size varies with angular
distance to mirror axis.}
\end{center}

Photons which fall in the insensitive areas of the camera are never
recorded. In the data analysis and detector simulation, the
unrecorded photons have been taken into account as part of the ray
tracing procedure. This is important in shower energy estimation and
tested using the stars crossing the field of view of the telescopes.
 When a bright star with a constant flux passes
through the FOV of the camera, the number of recorded photons varies
due to the different positions of the star on the camera. When the
star is near to the center of a PMT, most photons fall in the
sensitive areas of the camera, while when the star is near the
conjunction of two PMTs, most photons fall in the insensitive areas
of the camera, so the variations of the recorded photons from the
star on its track can be observed which is shown in the
Fig.~\ref{lightlost}. The variations are also effected by the
weather conditions, so several clear nights data is used to get an
average behavior of it. In the Fig.~\ref{lightlost}, the simulation
of the variations along the track of the star is also shown which is
consistent with the observed one. This demonstrates that the ray
tracing simulation correctly copes with the fraction of the
unrecorded photons in the insensitive areas of the camera.

\begin{center}
\includegraphics[width=9cm]{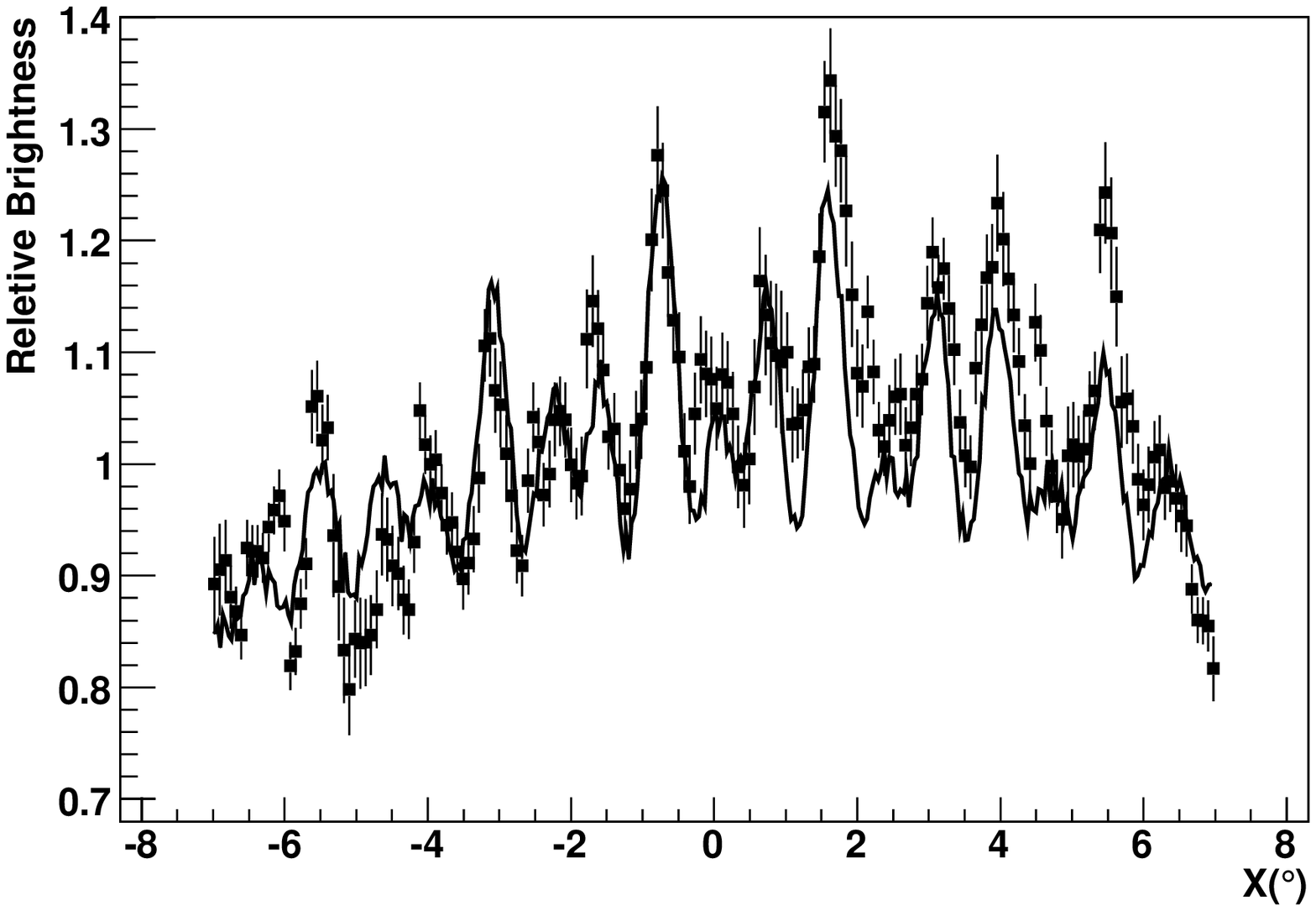}
\figcaption{\label{lightlost} The back squares show the variations
of the observed photons on the track of a star, while the black
curve shows the simulated one.
 }
\end{center}

\section{Conclusions}

Bright stars are used as a guider to calibrate the pointing
direction of each telescope and the geometry of the camera. The
pointing accuracy obtained through bright stars is better than
$0.05^{\circ}$. The long term stability of the pointing direction of
the telescope is also monitored by bright stars. Moreover, as point
sources, the bright stars are also used to study the spot shape. The
spot size becomes larger from the center to the edge of the camera.
Besides the spot size, the fraction of the unrecorded photons in the
insensitive areas between PMTs is compared with the simulated one,
and they are consistent with each other. The errors caused by this
effect are well understood and under control in the energy
reconstruction.

The pointing direction and the correction of the geometry and optics
of the telescopes are used in the simulation and data analysis to
improve the reconstructions of the energy and arrival direction of
the air shower.

\acknowledgments{ This work is supported by 100 Talents Programme of
The Chinese Academy of Sciences, Knowledge Innovation Program of The
Chinese Academy of Sciences ( H85451D0U2 ), National Natural Science
Foundation of China ( 10975145 ).}

\vspace{-2mm}
\centerline{\rule{80mm}{0.1pt}}

\end{multicols}
\clearpage

\end{CJK*}
\end{document}